\documentclass[doublespacing]{elsart1}
\usepackage{graphicx}
\journal{Nuclear Instruments \& Methods in Physics Research A,}
\begin{document}
\begin{frontmatter}
\title{\bf Near room temperature X-ray and   $\gamma$-ray spectroscopic detectors for future space experiments}
\author{ J. S. Yadav,}
\ead{ jsyadav@mailhost.tifr.res.in}
\author{S. Savitri and  J. P. Malkar}
\address{Tata Institute of Fundamental Research, Homi Bhabha road, Mumbai-400005}
\begin{abstract}
 New generation  Cadmium Telluride (CZT \& CdTe) solid state 
 detectors  
can provide   high quantum efficiency with reasonably good energy resolution 
and can  operate at near room temperature; an unique advantage for   
space experiments. We present here results of our study of small diode
detectors as well as large area pixel detectors. Our study is aimed at
developing near room temperature hard X-ray spectroscopy detectors for
ASTROSAT and other future Indian space science missions.
 We have studied a Si-PIN detector in the energy 
range 5 - 60 keV and
 CZT \& CdTe Schottky diode detectors in the energy
region 10  - 1330 keV.  Our results
suggest that the energy resolution is limited  by the thermal/electronic 
noise in the low energy region ($\le$ 60 keV) while it is affected by the 
charge transport properties of the detector in the higher energy region
 ($\ge$ 100 keV).
We have  studied large area (64 cm$^2$)  CZT pixel detectors with pixel size
close to the small CZT detector.  We have studied individual pixels as
well as  CZT detector as a whole (summed over all the 1024 pixels).
The CZT pixel detectors are single carrier detectors with peaking time 
$\le$ 1 $\mu$s. 
The energy resolution behaviour of the large area CZT
detector is similar to that of small  diode detectors in the low energy
region. The change in operating temperature from room temperature $\sim$ 
20$^o$C to $\sim$ 70$^o$C drastically affects both the energy resolution
as well as  the peak area due to a sharp rise in the thermal noise.
 These results suggest that the cooling of detector system 
will provide better energy resolution as well as  detection 
efficiency. Our results also show
that the crystal uniformity is another important issue in the case of 
large area CZT pixel detectors.  
\end{abstract}
\end{frontmatter}

\section{Introduction}

\begin{table*}[b]
\begin{center}
\centerline{Table 1: Properties of the semiconductor detectors} 
\begin{tabular}{lcccccc}
\hline\hline    
Semiconductor  & density& Z  & E$_{gap}$&$\epsilon$&X$_{0}$&$\Delta$E$_{intrinsic}$ \\ 
 & (g/cm$^3$) &    & (eV) & (eV) &(mm)& eV@100 keV\\
 \hline  
  Si   & 2.33  & 14   &   1.12   &  3.6   &  25&446 \\
  Ge   & 5.33  & 32   &   0.67   &  2.9   &3.3 &400  \\
  CdTe & 5.85  & 48, 52 & 1.44   &  4.43  & 1 &495 \\
  CZT  & 5.81  & 48, 30, 52 &1.6 &  4.6  & 1 &504 \\
\hline  
\end{tabular}
\end{center} 
 E$_{gap}$ : band gap energy, $\epsilon$ : ionization potential, X$_{0}$ : radiation length
\end{table*}

In hard X-ray spectroscopy (for astronomy or planetary mapping), the 
basic requirement is high sensitivity with high energy resolution which
can detect weak closely spaced lines above the background. The best
energy resolution among the diode detectors is obtained using  Ge diode 
detectors (E/{$\triangle$E}
$\sim$ 500), if they are cryogenically cooled to reduce the leakage
current and thermal noise. In space experiments, the detector 
requirements are high detection efficiency, high energy resolution and
operating temperature close to room temperature (does not need expensive
and cumbersome cryogenic cooling).
The ideal detector for space experiments should have high atomic number,
high density and a band gap much larger than that of Ge which can allow
operating temperature close to room temperature (Table 1 
lists some of the properties of the semiconductor detectors). 
In the last column, we give the intrinsic energy resolution calculated 
from statistical fluctuations in the number of electron-hole pairs created in
 photon interaction and assuming the fano factor f=0.1 for all 
the detectors [4,5].  Cadmium Telluride
 Cd$_{1-x}$Zn$_x$Te (CdTe and CZT) has been regarded as promising material for
hard X-ray studies.  Its high atomic number gives a high quantum efficiency and
a large band gap allows these detectors to operate at near room temperature.
 However the high defect density in these detectors
 results in considerable charge trapping. This leads to charge transport
properties far inferior to those of Si and Ge detectors.  The high defect 
density is also the main source of leakage current in these detectors.

Till recently, CZT was the best hard X-ray spectroscopy detector operating 
at near room temperature as it has  lower leakage current than the CdTe
detectors. The charge transport properties were the limitation
of CZT detectors: for electron 
${\mu_e}{\tau_e}{\sim}$ 1-2$\times$10$^{-3} cm^2/V$ and
for hole $\mu_h$$\tau_h$$\sim$ 1$\times$10$^{-5} cm^2/V$, 
where $\mu$ is the mobility
and $\tau$ is the life time of the charge carriers [6].  
This severely limits the 
effective depth of CZT detectors and hence  the energy range. The effective 
depth of 2 mm thick CZT detectors (investigated here) 
is commonly in the range of 0.5 to 1 mm.
The CdTe detectors show  superior charge transport properties: 
for electron ${\mu_e}{\tau_e}{\sim}$ 1-2$\times$10$^{-3} cm^2/V$ and
for hole $\mu_h$$\tau_h$$\sim$ 2$\times$10$^{-4} cm^2/V$.  The hole
lifetime in CdTe is typically 2$\times 10^{-6}$ sec, more than an order of 
magnitude larger than in CZT.  Thus,  a 1 mm thick CdTe
detector (investigated here) will have effective depth close to its 
physical depth. 
Recently developed CdTe Schottky diode 
detectors show further reduction in leakage current which permits the use 
of a much higher bias voltage, improving further the
transport properties and the effective depth.  
The trapping lengths ($\sim \mu_e \tau$ E where E is the electric field) 
for electrons and holes
in CZT are 2.7 cm and 0.01 cm respectively while these values in CdTe are
 13.2 cm and 0.8 cm for electrons and holes respectively (for E=400V).

\section{Experimental details}

\begin{table*}[b]
\begin{center}
\centerline{Table 2:  Some details of the diode detectors}
\begin{tabular}{llll}
\hline\hline
 Detector type  & CdTe Schottky diode &CZT & Si PIN diode\\
     Detector size  &  3x3 mm& 3x3 mm& 13 mm$^2$\\
     Detector thickness &  1 mm &2 mm & 300 $\mu$m\\
     Detector window    &  Be 250 $\mu$m & Be 250 $\mu$m & Be 25 $\mu$m\\
     Detector power[1]   &  400V @ 1 $\mu$A& 400V @ 1 $\mu$A & 100V @ 1$\mu$A\\
     Preamp. sensitivity&0.82 mV/keV &0.73 mV/keV &1 mV/keV\\
\hline
\end{tabular}
\end{center}
\end{table*}

\begin{figure}[t]
\centering
\includegraphics*[width=12cm,angle=0]{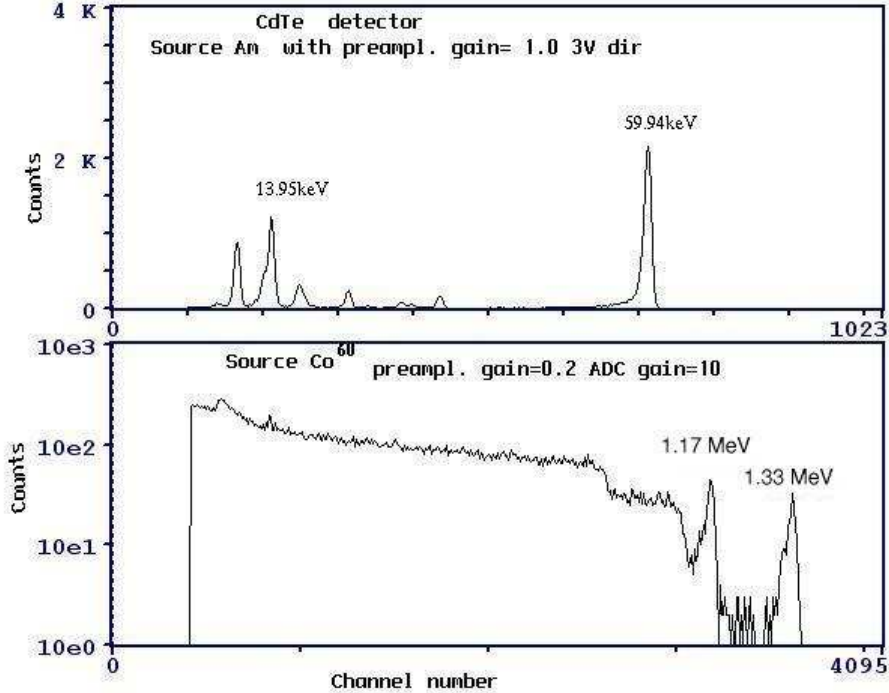}
\caption{The Am$^{241}$ spectrum measured with amplifier gain 1.0  
is shown in the top panel (acquiring time 200s) and Co$^{60}$ spectrum 
measured with amplifier gain 0.2, ADC coarse 10 and ADC fine 0.5 is 
shown in the bottom panel (acquiring time 30000s).}
\end{figure}

Main details of the diode detector systems which we have studied in this 
paper are
given in Table 2. All the detectors are from AMPTEK INC., USA. Specific power 
supply provides the main amplifier and the DC  high voltage needed to 
operate each detector system. The thin Si-PIN diode detector (300 $\mu$m
thick) is intended  to study soft X-rays below 30 keV while thick 
CZT and CdTe detectors are intended to study hard X-rays. The 
signal from the amplifier/power supply  can be connected directly to  a 
MCA (multichannel analyzer) or an oscilloscope.
We have used pocket MCAs (model 166 from GBS and  MCA 8000A from AMPTEK) 
connected to a PC or laptop 
for acquiring and storing the spectra. 
All the  detectors are  cooled using 
thermoelectric cooler to $\sim -30^o$C along with input FET transistor,
A250 charge sensitive preamplifier and the current feedback network [1]. A
temperature monitoring integrated circuit is placed  on the cooled substrate
to provide a direct reading of the temperature of the internal 
components.

The cooling of the detector reduces the leakage current considerably, thus
permitting the high positive bias voltage of 400V in case of CZT 
and CdTe detectors
which further improves the charge transport
properties. Recently developed CdTe Schottky diode detectors which are 
fabricated  using blocking (Schottky) diodes rather than the symmetric 
contact on CdTe show large reduction in leakage current [7].  
The charge transport properties in CdTe detector are  
much superior to those of  CZT, still the charge loss is not negligible.
The electron/hole pairs created  in the detector 
 near the back contact by radiation  result in fluctuations in charge 
collection time. As a result, the acquired spectra suffer from increased 
background counts and degraded energy resolution. To reduce these effects,
a rise time discrimination (RTD) circuit has been used in the  
power supply/amplifier with shaping time $\sim$ 3 $\mu$s.
When the RTD is ON, the shaped pulses  are internally gated  and only 
pulses corresponding to ``full charge collection'' events are allowed  to 
be  sent to  the MCA. In all our observations, the RTD is kept ON.

We have also studied large area CZT pixel detectors from IDEAS, Norway
and the details are given in section 4 along with the results.

\section{Observations and Results}

\begin{figure}[t]
\centering
\includegraphics*[width=9cm,angle=270]{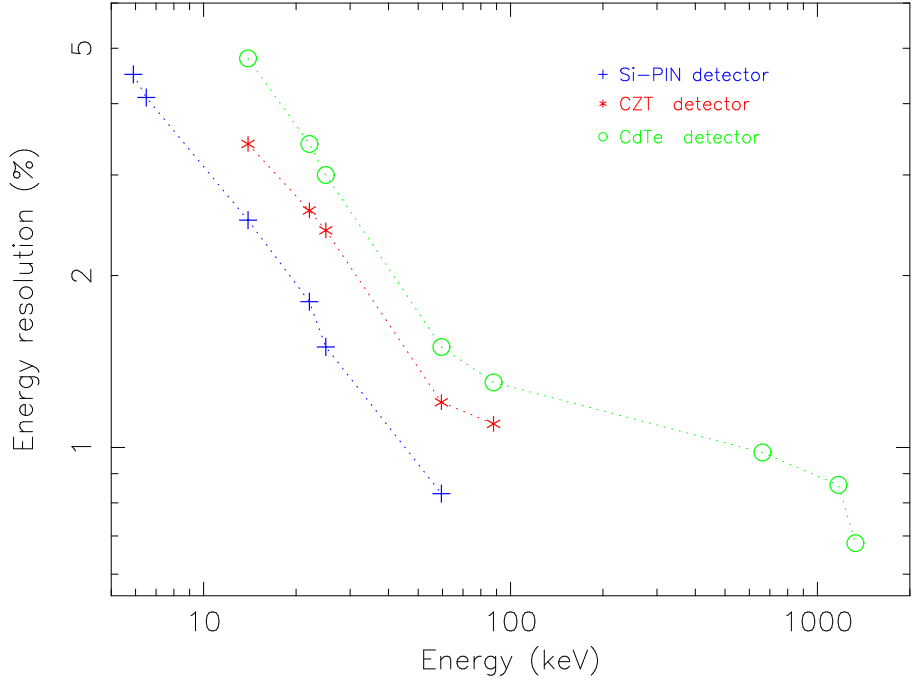}
\caption{The measured energy resolution (FWHM) for Si-PIN, CZT and CdTe 
detectors are shown as data points. The lines are to guide eyes.}
\end{figure}

In Figure 1, we show Am$^{241}$ spectrum in the top panel for 200 sec 
acquisition time and Co$^{60}$ spectrum in the bottom panel for 30000 sec
acquisition time using CdTe detector. All important energy lines are marked. 
These results
demonstrate the high efficiency of the CdTe detector in a wide spectral range
from 10 keV to 1330 keV  with good energy resolution. Our results of 
measured energy resolution vs energy for CdTe, CZT and Si-PIN detectors
are shown in Figure 2.   The energy resolution is defined as
the FWHM of a peak in \% of the energy.  The Si-PIN detector
provides better energy resolution than CdTe and CZT detectors in low
energy region. The quantum efficiency  of Si-PIN limits its energy
to 30 keV. We could also not detect the 662 keV line with the 2 mm thick
CZT  detector.  These results suggest high 
hole trapping and poor charge collection in CZT at higher energies 
and much superior charge 
transport properties of  CdTe detectors. This limits the energy range of
the CZT detector to the range between 10 keV  and a few hundred keV. 
However the CZT detector provides better 
energy resolution than the  CdTe detector suggesting lower 
thermal/electronic 
noise. These results also suggest that a thicker CZT detector (as 
used in the CZT array) will not increase much the energy range but only
affect adversely the energy resolution (see next section). We have also
studied the effects of electronic noise, high radiation flux and
detrapping by annealing.  Results are discussed in a separate 
publication [13].

\begin{figure}[t]
\centering
\includegraphics*[width=9cm,angle=270]{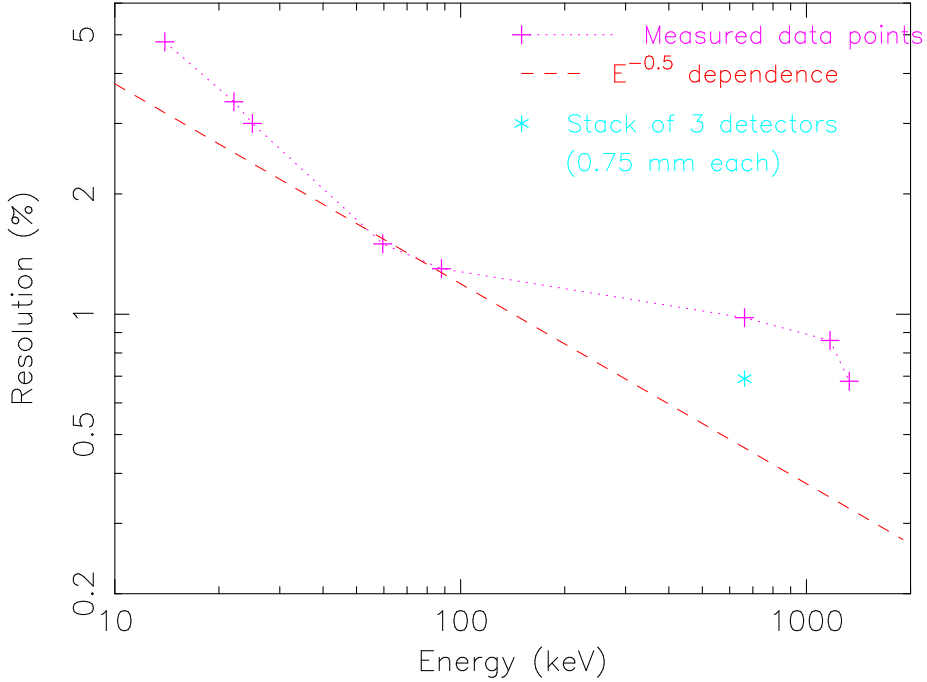}
\caption{The energy resolution (FWHM) of CdTe detector system with expected 
E$^{-0.5}$ dependence. The results of a stack CdTe detector is shown 
by star. For details see text.}
\end{figure}

Figure 3 shows the  measured energy resolution of the CdTe detector in the 
energy range 10-1330 keV (data points along the  dotted line). 
The energy resolution
is in the range 1-5 \% below 88 keV and is around 1\% in the wide 
energy range of 88 keV to 1330 keV. Due to the high atomic number, the 
photoelectric absorption is the main process up to 300 keV for CdTe as
compared to 150 keV for Ge. The 1 mm thick CdTe  has $\sim$ 64\% 
detection efficiency at  100 keV which reduces to  $\sim$ 4\% at 1000 keV.

\begin{figure}[t]
\centering
\includegraphics*[width=12cm,angle=0]{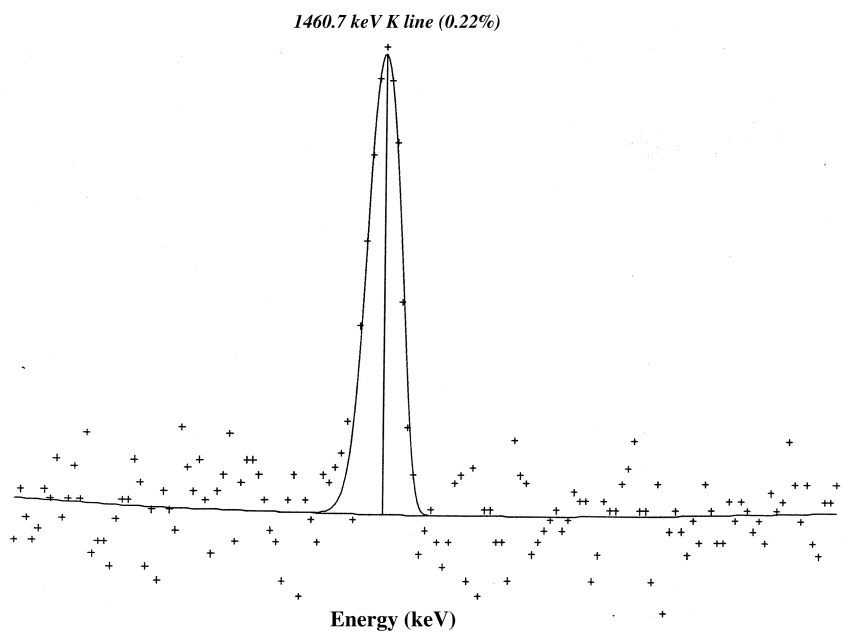}
\caption{The spectra of a neutron activated sample in the energy range 1420 -
1490 keV with High Purity Intrinsic Germanium detector system.}
\end{figure}

The background at 59.5 keV line in Am$^{241}$ spectrum and at 88 keV line
in Cd$^{109}$ spectrum is negligible suggesting very little contribution
from  thermal/electronic noise. These energies are still low and interactions
still occur close to the detector surface and one can reasonably assume
that the effect of charge collection loss is negligible at these energies. 
We fit
expected energy dependence of E$^{-0.5}$ in this energy range (dashed line)
and study the deviation of the measured energy resolution. The energy 
resolution is increasingly affected at lower energies by thermal/electronic 
noise below 60 keV while at higher energies ($>$  100 keV), the charge 
collection properties severely affect the energy resolution (operating 
temperature -30 $^o$C). At 1330 keV
the observed energy resolution is  0.68\% while expected resolution 
is  0.2\% (dashed line). For 662 keV line, measured energy resolution
is 0.98\% while expected energy resolution is 0.46\%. A stack of three
CdTe detectors with each of 0.75 mm thick (2.25 mm total thickness) provides
0.69\% energy resolution at 662 keV (shown by star in the Figure 3, 
AMPTEK Web-page). A stack
of thin CdTe detectors provides better charge collection properties and hence
improves  energy resolution. Thus a stack of thin CdTe detectors
can cover a wide energy range at  reasonably good energy
resolution [8]. 

  
In X-ray spectroscopy or in planetary mapping, one would desire to detect
weak lines above background which requires a high performance detector with 
very good energy resolution. It is known that the background continuum emitted 
from  the moon surface  has substantially larger flux  than the sum of all
the lines of interest.  It may be noted here that a large size Ge detector 
is being used in Mars Odyssey mission for mapping of the Mars. 
In Figure 4, we show the energy spectrum of a neutron-activated sample in the 
range  1430 keV -  1500 keV as measured by a Ge detector cryogenically 
cooled to 77 K [2].  The spectrum exhibits 
a strong K line at 1460.7 keV with 0.22\% energy resolution. 
This value is close to
the expected energy resolution in CdTe detector  as discussed before. 
We therefore conclude that the deterioration of energy resolution in 
thick CdTe is caused by  bad charge transport properties of the detector
material and a better energy resolution can be achieved by a stack of 
thin detectors.

%
%
%
%
\begin{figure}[t]
\centering
\includegraphics*[width=13cm,angle=0]{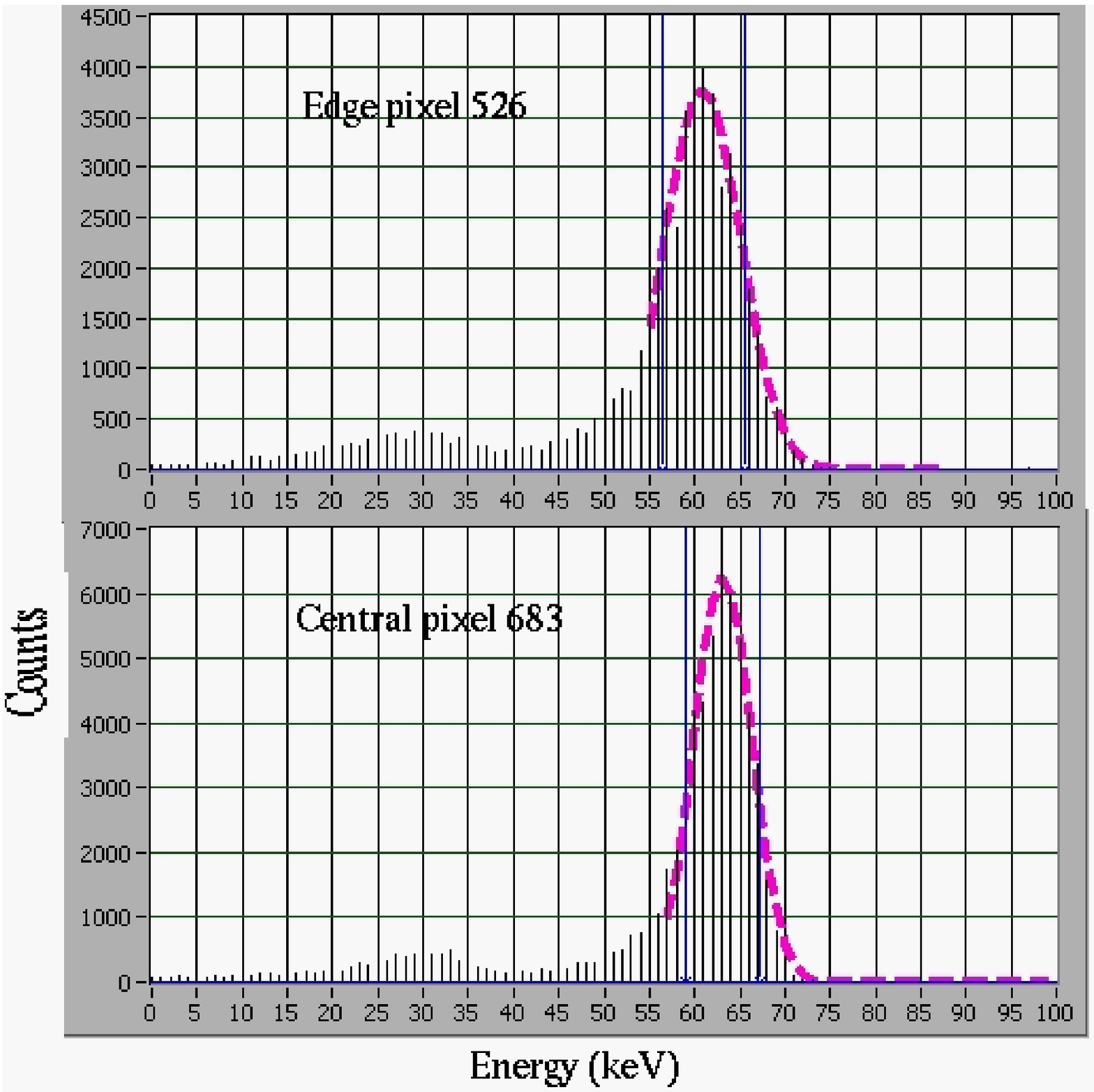}
\caption{The energy spectra from pixels 526 and 683 along with Gaussian fits.}
\end{figure}

\begin{figure}[t]
\centering
\includegraphics*[width=10cm,angle=270]{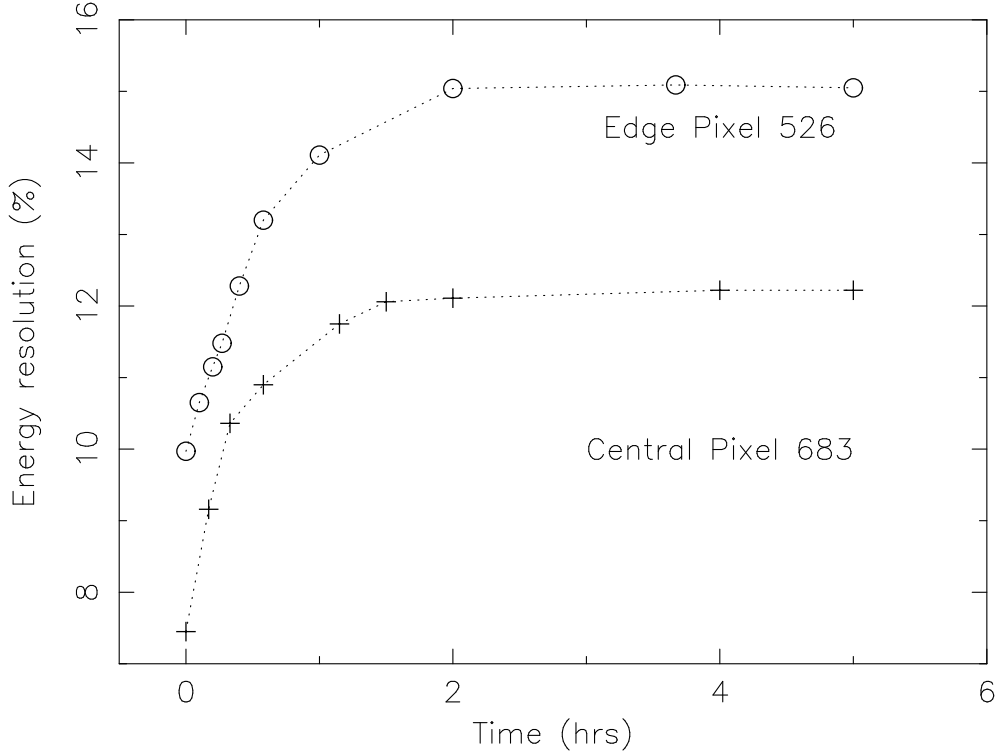}
\caption{Change in the energy resolution ($\sigma$) for individual pixels  
with time as detector temperature rises. Lines are to guide eyes.}
\end{figure}

\begin{figure}[t]
\centering
\includegraphics*[width=10cm,angle=270]{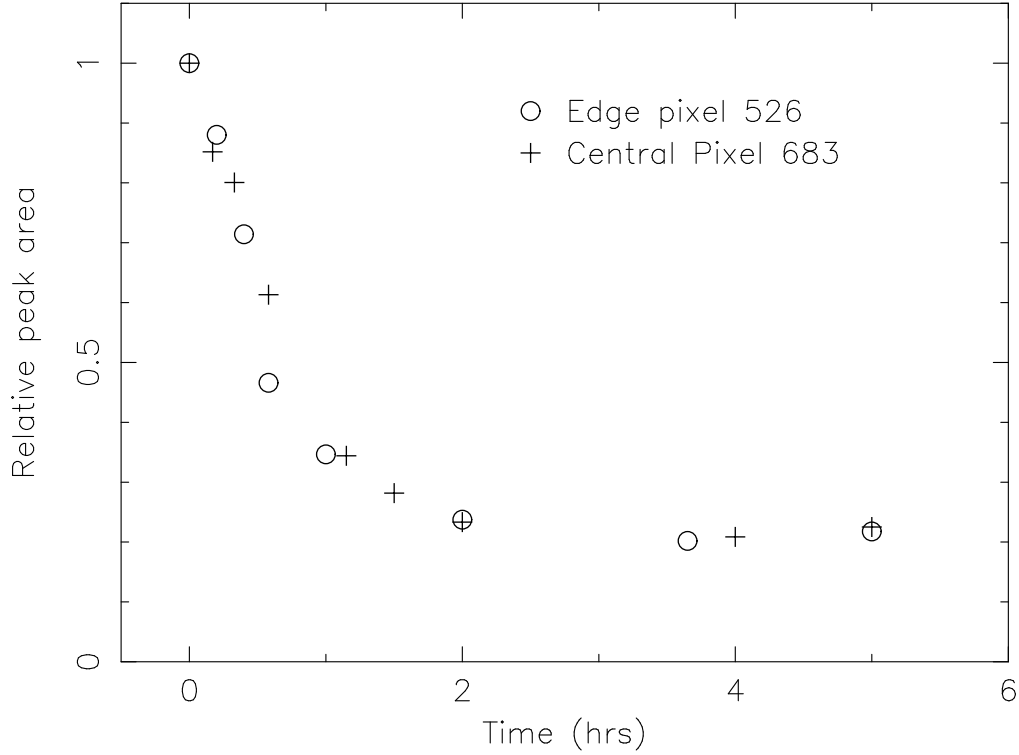}
\caption{Change in the relative peak area for individual pixels with time 
as detector temperature rises.}
\end{figure}

\begin{figure}[t]
\centering
\includegraphics*[width=10cm,angle=270]{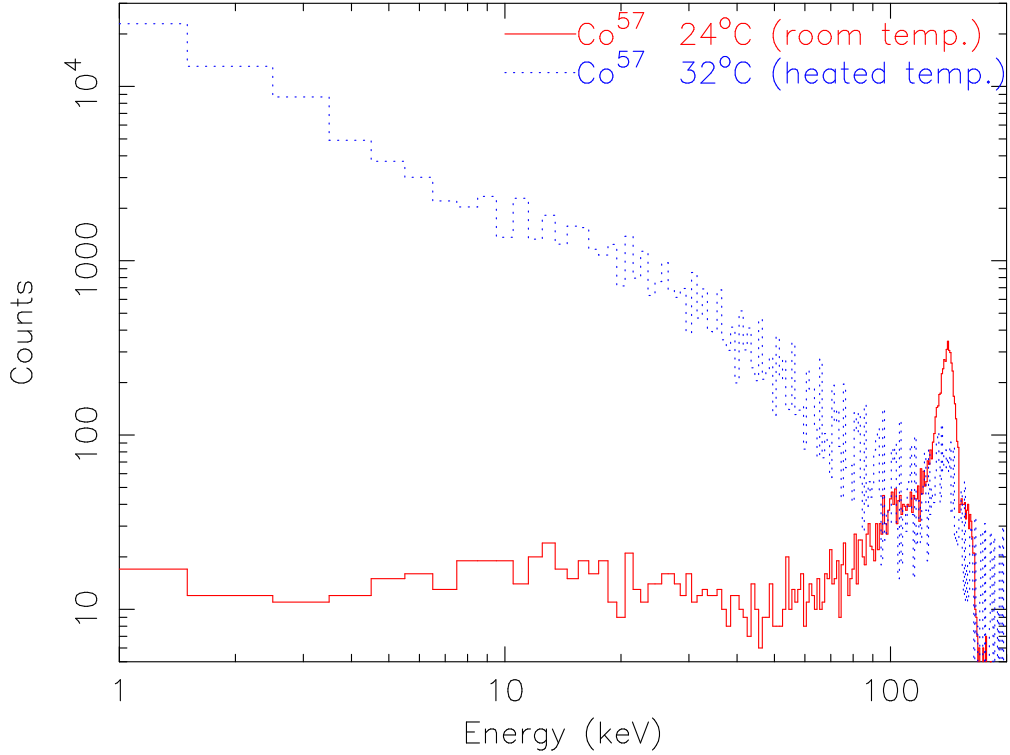}
\caption{The spectra of Co$^{57}$ taken at 24$^o$C room temperature and 
32$^o$C heated temperature after two hours operation.  The spectra are 
summed over all the pixels.}
\end{figure}

\begin{figure}[t]
\centering
\includegraphics*[width=10cm,angle=270]{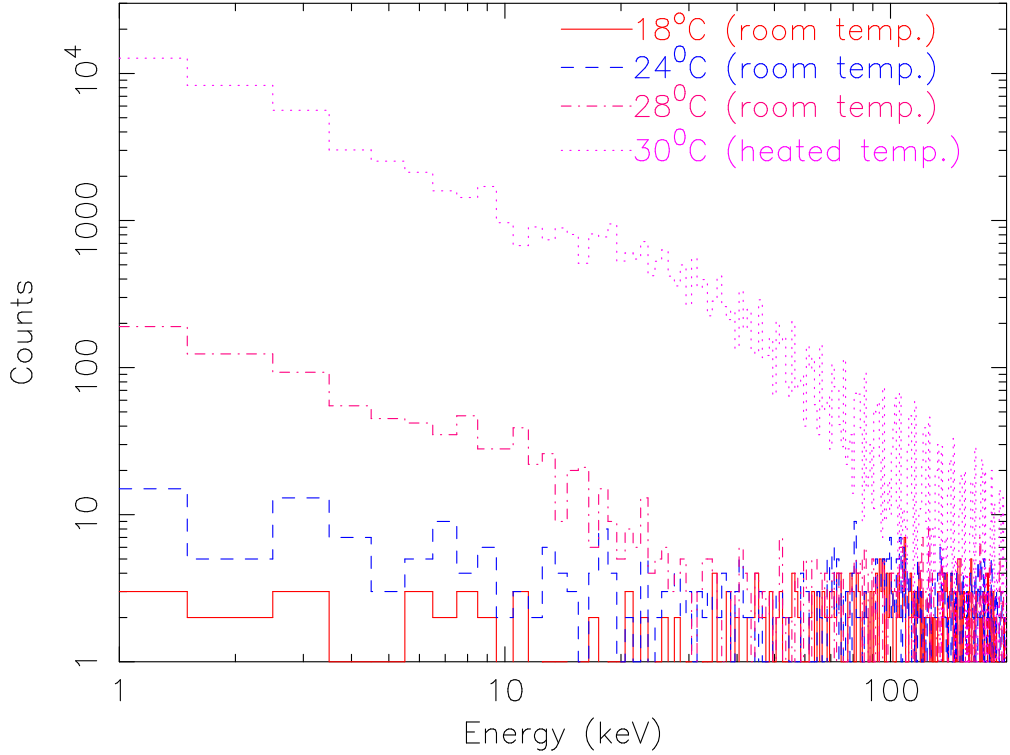}
\caption{The background spectrum (summed over all the 1048 pixels) taken
at room temperature 18$^o$C, 24$^o$C, 28$^o$C and at heated temperature
30$^o$C (after two hour operation).}
\end{figure}
\begin{figure}[t]
\centering
\includegraphics*[width=11cm,angle=270]{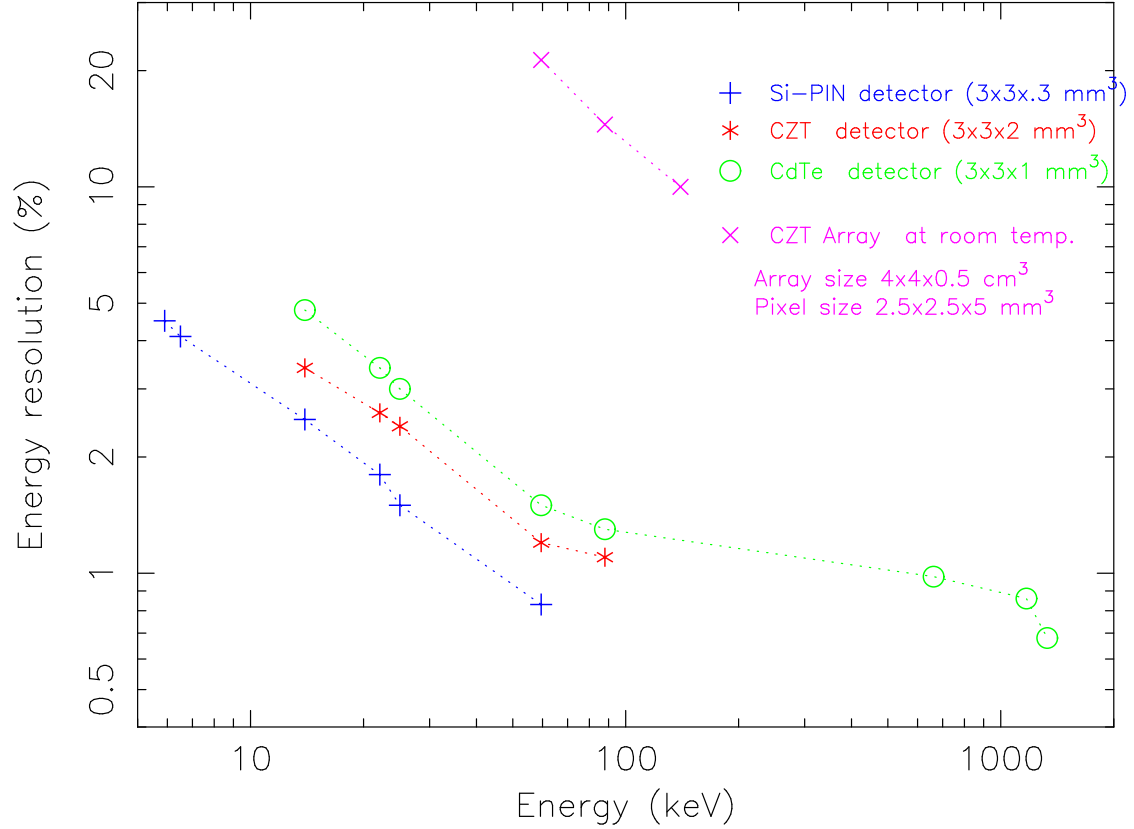}
\caption{The energy resolution (FWHM) of the  large area CZT detector  
  along with energy resolution data of  Si-PIN, CZT and 
CdTe small diode detectors.  The lines are to guide eyes.}
\end{figure}

\section{Test with large CZT array}
	Initially the large area pixel detectors were planned as  CZT 
detectors  as this material
provides by an order of magnitude lower leakage current [6].  The small
pixel size is one way to reduce further the leakage current as well as 
it can provide imaging.  Handling
a large number (several thousands) of amplifiers and related front-end
electronics is one of the critical R\&D areas in the development
of large area CZT detectors. Normally, this is achieved by designing
Application Specific Integrated Circuits (ASIC). We have studied large area 
CZT array systems provided by Ideas, Norway
(with XAIM3.2 ASIC readout)  which use CZT crystals
developed by IMARAD Imaging System, Israel. Each module uses 4$\times$4 cm$^2$ 
CZT crystal with 0.5 cm depth (crystals are grown using the modified 
Bridgeman method). 
The pixel size is 2.46$\times$2.46 mm$^2$.
The total pixels are 256 per module  (16$\times$16) and each module 
has its own ASICS
(2 for each module) [9]. It uses Gold/Gold contacts and a floating 
guard band of the adhesive tape with a metallic strip deposited near 
one end [12]. It is known that the IMARAD CZT material is slightly N type
and a negative bias of -600V is used. The default peaking time is 
0.5 $\mu$s. The leakage current is in the range of 10-30 nA.
We can operate four such modules in our current
system which has arrangement to cool the detectors during operation by
a liquid coolant. The detector module is supposed to provide less than
10 \% energy resolution at 140 keV with more than 40 \% photo peak efficiency.

We show in Figure 5 the spectra of Co$^{60}$ for 10 sec accumulation time from
an  edge pixel  and a central pixel (pixel numbers 526 and 683 
respectively)  at 24$^0$C. The thick dotted
lines show the Gaussian fits to high energy region of the peaks (excluding
the tailing part at  low energy). The central pixels usually provide better
energy resolution than  the edge pixels.

	Each detector module/ASICS system consumes $\sim$ 0.82 W power.
If cooling is not used, the dissipation of heat will  raise the 
temperature of the system. In this work, we study the detector system without
cooling.
We have studied the effect of rising temperature
on the energy resolution. The energy resolution is given as standard
deviation (FWHM = 2.35 $\sigma$) in \% of peak position. We have used a thermal
sensor  to measure the temperature of 
detector/ASICS system. A special thermal grease is used for thermal contact
between the thermal sensor and the cooling finger attached to the 
detector/ASICS system.
The temperature rises from 24$^o$C at the start and
stabilizes at $\sim$ 32$^o$C in about two hours time (here 32$^o$C is a 
lower limit on the temperature of  the detector system which may be higher
due to the temperature gradient).   We assume that the detector system 
is initially  in equilibrium 
with room temperature and any quick measurements within 10-20 sec are
assigned to the room temperature. For longer durations, the detector 
heats up and a temperature gradient appears.
We will refer to it as  ``heated temperature''.
Later in this section we will show that the actual temperature of the
detector may be as high as 70$^o$C when our thermal sensor attached to
the cooling finger shows around 30$^o$C.  The result of heating   
is shown in Figure 6.
The energy resolution degrades during the first two hours. 
Initially the change in energy 
resolution is faster for the central pixel (\# 683) than  for the 
edge pixel (\# 526) as expected. This reveals a spatially inhomogeneous
 rise of thermal noise and background.  The 
energy resolution degrades by about 4\%. We have also studied the effect of 
rising temperature  on the  peak area. The peak area  is
calculated   within $\pm 3 \sigma$ assuming a Gaussian peak  shape
based on the  measured $\sigma$ and
peak height.  The results are shown in Figure 7. For both the pixels,
edge and central, the change in the peak area coincides, suggesting a 
global effect. It is most likely the dead time/multi hit due to sharp
rise in the thermal noise which  will be clear as we discuss our 
results of background at different temperature in Figures 8 and 9.

We show Co$^{57}$ spectra for 10 sec accumulation time in Figure 8  
taken at room  
temperature  24$^o$C and at heated temperature 32$^o$C after two hours 
of operation. Both the spectra are summed over 
all the 1024 pixels. The energy calibration is based on single energy
of 60 keV (Am$^{241}$). The 122 keV peak is clear in spectrum taken at 
room temperature  24$^o$C while this peak is hardly distinguishable  
in spectrum taken at heated temperature 32$^o$C. 
In two hours of operation,
there is a sharp rise in the thermal noise/background 
counts in the  low energy region. This is consistent with
our explanations of  Figures 6 \& 7.

To prove further we study background spectra at different temperature.
In Figure 9, we show the background spectra (summed over all the 1024 pixels)
taken at room temperature 18$^o$C, 24$^o$C, 28$^o$C and at heated   
temperature 30$^o$C (after two hour operation). From the shape of the  
background spectrum at 18$^o$C, it  may be safely assumed that it is
mostly of non-thermal origin with no significant  contribution from
thermal noise. The background spectrum taken at 28$^o$C (room temp.) shows
significant thermal noise contribution below 30 keV while at higher
energies above 30 keV, the background spectrum is similar to one observed
at 18$^o$C (IDEAS has suggested operating temperature range 18-28$^o$C
for our detector system in the energy range 30-200 keV). The population of 
conduction electrons for a semiconductor at temperature T is given by
 \[ N_{cb} = A e^{-E_g/2kT} \]  
    A is a constant (for the temperature range discussed here), E$_g$ is 
band gap energy (1.6 eV) and k is Boltzmann constant. We have used background
data at 28$^o$C room temperature and at 30$^o$C heated temperature (Figure 9).
The rise in background is by a factor $\sim$ 50 in the energy range 1-10 keV.
It gives the real temperature of the detector is $\sim$ 70$^o$C after two
hours of operation when our temperature sensor thermally attached to the
cooling finger shows $\sim$ 30$^o$C (the final equilibrium temperature
after two hours of operation also depends on initial room temperature).
These results show a raised temperature. A rise
of 50$^o$C (from room temperature $\sim$ 20$^o$C to $\sim$ 70$^o$C) affects
both the energy resolution and the peak area. It is important to use 
ASICS with lower power or use cooling to keep background low when long 
duration operations are needed.

\begin{figure}[t]
\centering
\includegraphics*[width=10cm,angle=270]{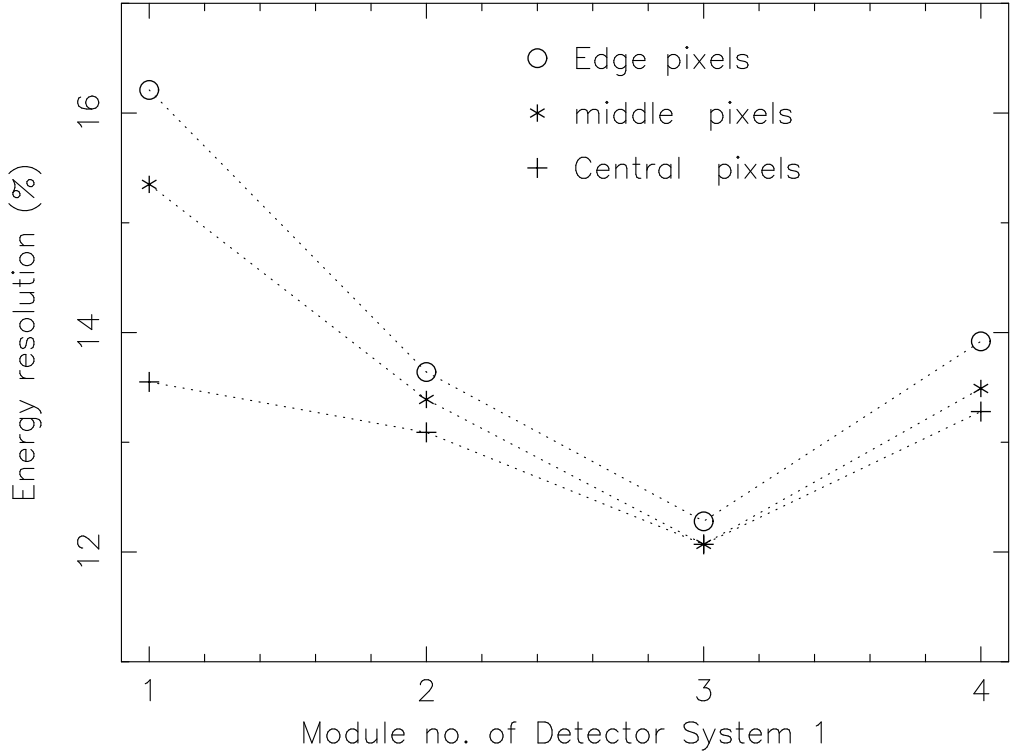}
\caption{The average energy resolution ($\sigma$) of central pixels, middle 
pixels, and edge pixels of four different detector modules. The lines are 
to guide eyes.}
\end{figure}

We have tested energy resolution of this system at  
 room temperature. During the test at room temperature
we have used four detector modules (64 cm$^2$ area with 1024 pixels).
The results are shown
in Figure 10 along with results of  small area diode  detectors. 
The energy resolution behaviour of the large area CZT pixel detector 
is similar to the energy resolution  
behaviour of the small area diode detectors in the low energy 
($\le$ 60 keV).
The large area CZT pixel detector is likely to have same  
behaviour in the 
high energy region (above 100 keV) as (1) it uses fast peaking time
of $\sim$ 0.5 $\mu$s, and (2) the large area CZT pixel detectors are
mostly single carrier detectors.
At room temperature (24$^o$C),
 the energy resolution of CZT array (summed over all pixels) is degraded  
by a factor of 18 compared to  that of a small CZT detector, which is 
only 2 mm thick and was     operated at -30$^o$C.
This factor rises to above  24 at $\sim$ 70$^o$C (after two hours of 
operation). 
Even the best pixels exhibit a factor of 15 at 24$^o$C.
It is known that changing operating temperature from 20$^o$C to -20$^o$C
results a drop in the leakage current by almost two order of magnitude [10].
Here, one can identify  two main factors which are likely to degrade the
energy resolution 
(1) the CZT array is being operated at room temperature while the small
CZT detector is operated at -30$^o$ and, 
(2) the CZT array is thicker (5 mm thick) than the small CZT detector 
which would result more  leakage current due to larger number of
defect/trapping sites.  Therefore, thiner CZT array
would  also be expected  to improve the energy resolution.
Both these factors will reduce  the leakage current
and hence improve  the energy resolution.  One needs to explore the low
operating temperature range for CZT array detectors as there are reports
that the IMARAD standard and Au detectors stop functioning around
-20$^o$C [11].
Recent results of the Schottky contact and the 
guard ring which show a drastic drop in leakage current (more than 
three order of magnitude) will bring further improvement in the energy  
resolution [8].

\begin{figure}[t]
\centering
\includegraphics*[width=10cm,angle=270]{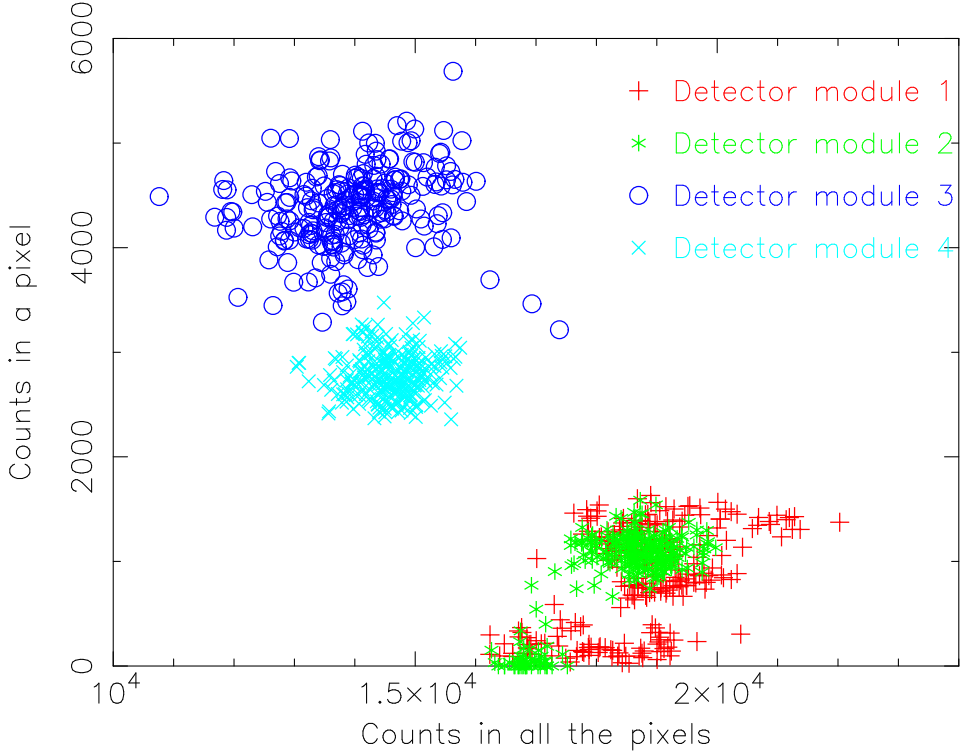}
\caption{Total counts in the spectrum from a pixel which is illuminated  
with collimated Am$^{241}$ (the beam size is less than the pixel size)    
vs the counts in the spectrum summed over all the 1048 pixels.}
\end{figure}

Another important issue with large area CZT pixel array detectors is the
uniformity of the crystal. We have measured  the energy resolution 
of individual pixels (central pixels, middle pixels and edge pixels) of 
the four CZT detector modules operating at 70$^o$C (after 2 hours of 
operation) and results are shown 
in Figure 11. For central regions, $\sigma$ is averaged over 10 pixels
while it is averaged over 20 pixels for middle and outer pixels (bad pixels
are excluded from this study).
On an average, central pixels show better energy resolution 
than the edge pixels which is in agreement with previous studies [12].
The detector modules 2, 3 and 4 show typical
variation in the energy resolution but detector module 1 shows large 
variation.  Our further study of this detector module shows large
patches  of bad pixels which show bad detector properties
suggesting non-uniform detector material. In Figure 12, the counts in 
individual pixel vs the counts summed over all the 1024 pixels are 
plotted for these four detector modules. A collimated Am$^{241}$
source was used to irradiate each pixel. 
The beam size was  less than the 
pixel size. This study includes all the pixels (including bad pixels).
The detector modules 3 and 4 show good  efficiency of individual pixels
while detector modules 1 and 2 show 
that many pixels have poor performance with low detection efficiency
(close to 0) in many cases.
These results suggest cross talk  between pixels as the counts
summed over all the 1024 pixels increase. Both of these suggest
bad crystal detection properties as well as non-uniformity of the 
detector material. It may be noted here that the energy resolution
of individual pixel improves as the counts in the individual pixel
increase (not shown).

\section{Conclusions}
New generation  Cadmium Telluride Cd$_{1-x}$Zn$_x$Te (CdTe and CZT) detectors provide great advantage of compactness and operating temperature close to
room temperature for future space experiments  due to their higher atomic 
number and higher band gap.   The thermal/electronic noise affects the energy 
 resolution in low energy while bad charge transport  properties affect 
 the energy resolution in
 higher energy range ($>$100 keV).  In case of large area CZT pixel 
 detectors, the energy resolution behaviour is similar to that of the
 small diode detectors in the low energy region.
 In the space, low ambient temperature will
 reduce the problem of thermal/electronic noise.
 The large area CZT pixel detectors can provide good energy resolution as
 well as  high quantum efficiency while operating at 
the  ambient temperature in the space and  hold bright prospect for future 
space experiments. 
It is capable of  detecting  most of the  lines of interest in the case of moon
$\gamma$-ray mapping [2]. For better energy resolution and better peak
detection efficiency,
a basic requirement in low count rate astronomy and planetary science, our
results suggest the following:
1. use of low power ASICS, 2. decouple detector and ASICS (as most of the 
heat is produced in ASICS), and 3. use cooling to operate  
detector at low temperature.
The INTEGRAL and SWIFT space missions are already using these new generation
detectors to achieve their scientific goals [14,15].
Future developments in the crystal technology of 
 compound semiconductors will reduce the gap in the energy
resolution achieved with the cryogenically cooled Ge detector and CdTe 
detectors.

The Indian Multi-wavelength Satellite, ASTROSAT, has been planned with 
broad-band X-ray spectroscopy as one of its major objectives. One of 
its experiments is a large area  (1024 cm$^2$) CZT detector array. It will
have a total of 16384 pixels and  the pixel  size  will be  
2.5 x 2.5 mm$^2$ (5 mm depth) [3].  Recent
progress in CdTe has shown  superior charge transport properties and the
effective depth close to its geometric  depth. 
A stack of thin  CdTe detectors can
cover  a wide energy range from 10 keV to  a few  MeV with reasonably
 good energy resolution.
If we replace pixel material from CZT to CdTe, the same configuration 
will provide enormous advantage in the same energy range.  
These detectors will have very good efficiency close to 100\%
up to about 200 keV due to the superior charge transport properties of 
CdTe over CZT with similar or better energy resolution. Such a wide energy
range (say 10 keV to 1 MeV; higher energy is limited by the detector size 
as well as 
the photon flux) will have enormous advantage in the  study of non-thermal 
emission in astronomy.  

\section*{Acknowledgments}
We thank anonymous referee for constructive and very useful comments.
We thank  Prof. A. R. Rao  and Prof. P. C. Agrawal for support and 
encouragement.  We  thank Mr.  Parag Shah for technical help. We also 
thank Mrs S. Prabhudesai for help during these observations.

\end{document}